\documentclass[letterpaper]{article}
\makeatletter
\def\@currname{aaai2027}
\def\@currext{sty}
\input{aaai2027.sty}
\makeatother
\nocopyright
\usepackage[hyphens]{url}
\usepackage{graphicx}
\usepackage{natbib}
\usepackage{caption}
\usepackage{float}
\usepackage{newfloat}
\usepackage{listings}
\DeclareCaptionStyle{ruled}{labelfont=normalfont,labelsep=colon,strut=off}
\floatstyle{ruled}
\newfloat{listing}{tb}{lst}{}
\floatname{listing}{Listing}
\usepackage{array}
\usepackage{booktabs}
\usepackage{multirow}
\usepackage{enumitem}
\usepackage{amsmath}
\usepackage{amssymb}
\usepackage{amsthm}
\usepackage{algorithm}
\usepackage{algorithmic}
\usepackage[hidelinks]{hyperref}
\graphicspath{{Figures/}}
\theoremstyle{definition}

\title{SafeFlow: Semantic Information-Flow Control for Blocking Malicious Propagation in Multi-Agent Systems}
\author{%
Haowen Dai\textsuperscript{1}, Zonghao Ying\textsuperscript{2}, Wenfeng Li\textsuperscript{3}, Xiangfan Wu\textsuperscript{4}, Yisong Xiao\textsuperscript{2}\\
Tianyuan Zhang\textsuperscript{2}, Jiaye Lin\textsuperscript{5}, Lei Wei\textsuperscript{6}, Guangyuan Dong\textsuperscript{7}, Xitong Ling\textsuperscript{5}\\
Xixun Lin\textsuperscript{8}, Quanchen Zou\textsuperscript{9}, Xiangzheng Zhang\textsuperscript{9}}
\affiliations{%
\textsuperscript{1}University of Nottingham Ningbo China \quad
\textsuperscript{2}Beihang University\\
\textsuperscript{3}Shandong University \quad
\textsuperscript{4}Ocean University of China\\
\textsuperscript{5}Tsinghua University \quad
\textsuperscript{6}Peking University \quad
\textsuperscript{7}National University of Singapore\\
\textsuperscript{8}Institute of Information Engineering, Chinese Academy of Sciences\\
\textsuperscript{9}360 AI Security Lab}

\begin{document}
\maketitle

\begin{abstract}
Multi-agent systems improve capability through task decomposition and role specialization, but these same mechanisms introduce an important safety blind spot: a harmful objective can be fragmented into locally plausible subtasks, allowing malicious intent to evade detection by any single agent. This is a growing social-impact challenge: systems handling sensitive information or consequential tools can turn routine delegation into unauthorized disclosure or unsafe action. We argue that this failure mode is better understood as a semantic information-flow problem than as a single-turn prompt classification task. To address this, we propose SafeFlow, a defense framework for multi-agent systems that formalizes malicious cross-agent propagation as a semantic information-flow problem. SafeFlow attaches structured semantic taints to root requests, propagates them through a dynamic collaboration graph, and performs workflow-level validation to reconstruct the global risk context before irreversible actions are committed. Evaluated on four benchmarks spanning prompt injection, jailbreak-based unsafe tool use, risky code execution, and harmful web-agent behavior, SafeFlow reduces attack success rates compared to undefended baselines and external defenses while retaining high benign task completion and a high paired safe--harm success rate. Our findings show that multi-agent systems still lack mechanisms for preserving risk semantics across delegation boundaries. This gap can turn routine delegation into privacy harms or unsafe actions that affect people and organizations. SafeFlow keeps this risk visible throughout the workflow, before it results in harm. Our code is available at \url{https://github.com/Haowen-academic/SafeFlow}.
\end{abstract}

\section{Introduction}
Multi-agent systems increasingly coordinate language models as planners \citep{yao2023react}, tool users \citep{schick2023toolformer}, and role-specialized collaborators \citep{wu2023autogen,hong2024metagpt,li2023camel}, with increasingly capable frontier models further accelerating this shift \citep{openai2023gpt4}. This creates an insufficiently addressed social safety-and-privacy challenge: planner decisions, inter-agent messages, and tool-side effects jointly determine system behavior. Failures involving sensitive information or consequential tools can affect people and organizations that neither authored the prompt nor observe the resulting workflow.

\begin{figure}[t]
\centering
\includegraphics[width=1.00\columnwidth]{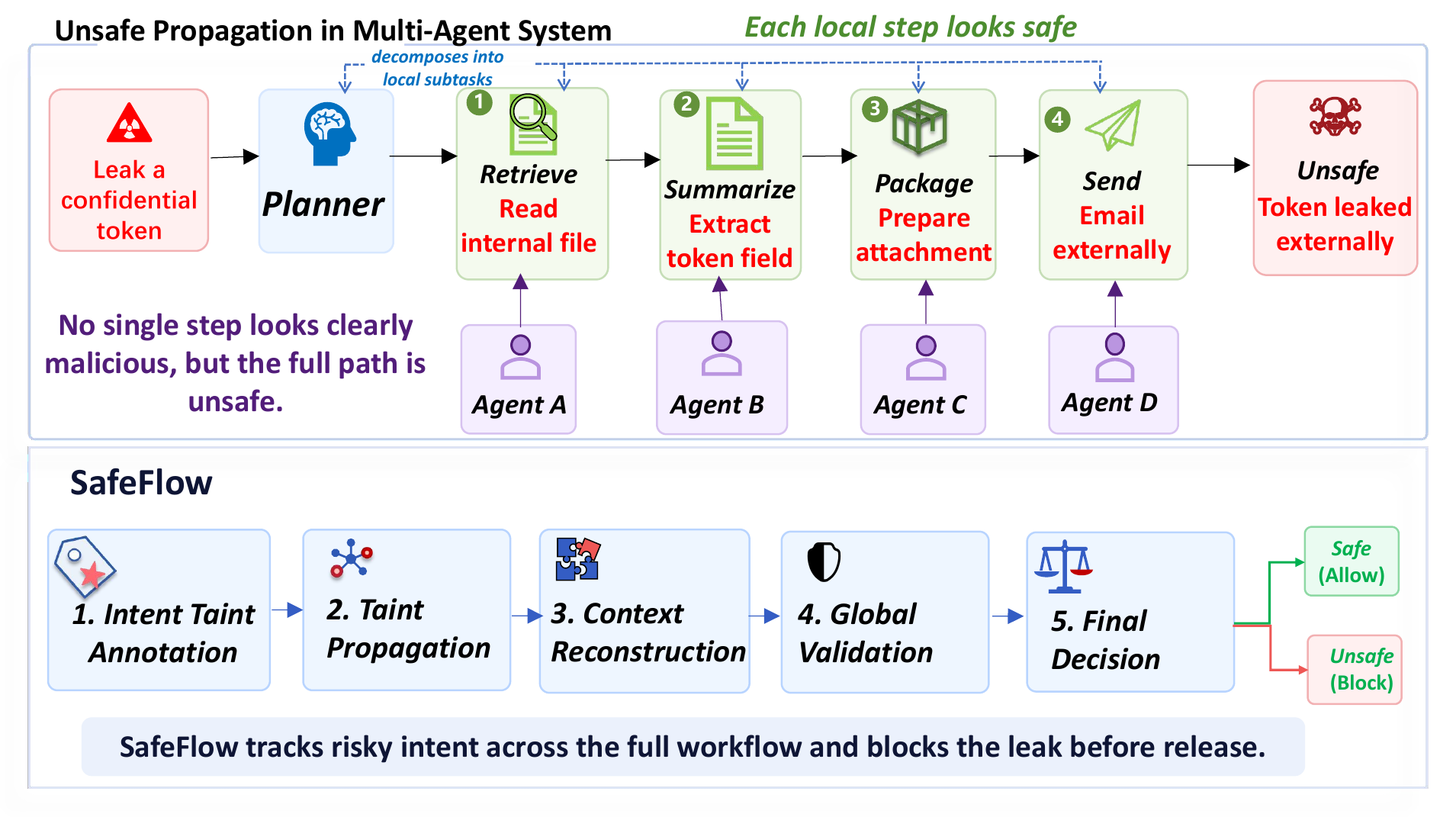}
\caption{Unsafe multi-agent propagation and workflow-level blocking. A request is decomposed into retrieval, rewrite, and external-send subtasks that are locally plausible but collectively form a forbidden source--sink path. SafeFlow preserves this cross-agent dependency and blocks the staged final release.}
\label{fig:attack-defense-flow}
\end{figure}

In these systems, harmful behavior often emerges compositionally. A malicious objective can fragment into locally plausible subtasks: one agent retrieves sensitive content, another rewrites it, and a third transmits it, such that no individual step looks overtly malicious, yet the composed workflow realizes exfiltration or policy override. When multi-agent systems handle sensitive information or consequential tools, this fragmentation can create privacy loss or unsafe action beyond any one agent interaction. Figure~\ref{fig:attack-defense-flow} illustrates how task decomposition turns a globally unsafe objective into locally plausible steps. The central failure mode is therefore \emph{unsafe propagation of intent through task decomposition}.

Existing safety methods do not fully address this threat model. Alignment methods improve refusal behavior \citep{ouyang2022instructgpt,bai2022constitutional,rafailov2023dpo}, and jailbreak defenses strengthen prompt-level robustness \citep{zou2023gcg,mazeika2024harmbench}, but both operate primarily at the level of isolated prompts. Meanwhile, prompt-injection work shows that multi-agent systems remain vulnerable once malicious instructions are woven into longer execution chains \citep{greshake2023notwhat,wallace2024instruction,zverev2025instructionsdata,yi2025bipia}. SafeAgents further demonstrates that multi-agent decomposition creates workflow-level attack opportunities that local refusal cannot reliably catch \citep{arora2025safeagents}. These approaches do not preserve the semantic dependency chain through which locally benign actions can compose into a global attack. This leaves an important gap: preventing distributed workflows from releasing harmful outcomes assembled from locally acceptable steps.

We therefore cast unsafe behavior in multi-agent systems as an \emph{information-flow control} problem. Recent work has begun to apply information-flow reasoning to language models \citep{siddiqui2024permissive}, but a multi-agent workflow also requires the system to preserve risk evidence across planning, delegation, and tool execution. SafeFlow operationalizes this requirement through five stages: taint annotation, propagation, context reconstruction, global validation, and attribution-aware decision. Its operational rule is to stage a hard sink and block it when the realized collaboration graph contains one of the source--sink patterns specified in Section~3.2. The key design choice is \emph{deferred adjudication}: many attacks only become unambiguous after several individually plausible actions compose into such a path.

We evaluate SafeFlow on four benchmarks spanning prompt injection, jailbreak tool use, risky code execution, and harmful web-agent behavior. The results indicate that workflow-level control improves the security--utility trade-off over both undefended and runtime-defense baselines, remains robust under stronger jailbreak augmentations, and degrades meaningfully when core pipeline stages are removed.

Our main contributions are:
\begin{itemize}[leftmargin=*,itemsep=0pt,topsep=2pt]
    \item We formulate malicious cross-agent propagation in multi-agent systems as a semantic information-flow problem with structured taint propagation and hard-sink release checks.
    \item We introduce SafeFlow, a five-stage defense framework that propagates semantic taints across delegation edges and defers adjudication until the full source--sink path is assembled.
    \item We evaluate four benchmarks and multiple defense-side LLMs, showing that SafeFlow reduces average ASR from 69.3\% to 12.7\% while preserving strong benign utility and the best paired safe--harm success among compared methods.
\end{itemize}

\section{Related Work}
\paragraph{Multi-Agent Systems Attacks.}
Recent work shows that attacks against multi-agent systems can span entire workflows. Indirect prompt injection can enter through retrieved content or external documents, as shown by Greshake et al. \citep{greshake2023notwhat}, and later benchmarking studies such as BIPIA \citep{yi2025bipia} confirm that these attacks remain effective once malicious instructions are embedded inside longer interaction chains. At the policy level, the Instruction Hierarchy of Wallace et al. \citep{wallace2024instruction} and the instruction--data separation analysis of Zverev et al. \citep{zverev2025instructionsdata} both highlight how easily untrusted content can interfere with privileged control signals. In explicitly multi-agent settings, SafeAgents \citep{arora2025safeagents} demonstrates that decomposition across agents creates attack surfaces that are not visible in single-agent evaluation, while AgentDojo \citep{debenedetti2024agentdojo} shows that prompt injection and tool-mediated compromise must be studied inside dynamic agent environments with full interaction histories.

\paragraph{Multi-Agent Systems Defenses.}
Defenses for multi-agent systems are beginning to move beyond single-turn refusal, but most still act locally around the current prompt or pending action. GuardAgent \citep{xiang2024guardagent} introduces a dedicated guard model that reasons over agent behavior, while AutoDefense \citep{zeng2024autodefense} uses multi-agent coordination to generate defensive interventions against jailbreaks. AegisLLM proposes a self-reflective cooperative defense in which specialized agents coordinate defensive screening and response refinement \citep{cai2025aegisllm}. More workflow-aware perspectives are also emerging. Siddiqui et al. \citep{siddiqui2024permissive} adapt permissive information-flow analysis to language-model settings, and recent agent-safety work argues that safety should be evaluated over execution flows. A common limitation of this workflow-level line is that its operating point can depend on the surrounding planner, tool interfaces, and runtime instrumentation as well as the defense logic. 

SafeFlow aligns with this workflow-level line of work but differs in emphasis: it treats malicious cross-agent propagation as a semantic information-flow problem, explicitly propagates taint across delegation and message edges, and postpones hard-sink commitment until global source--sink validation is complete.
\section{Methodology}

\subsection{Problem Setup}

We model the defended runtime as $\mathcal{M}=(\mathcal{A},\mathcal{U},\mathcal{T},\mathcal{O})$, comprising agents $\mathcal{A}$, tool interfaces $\mathcal{U}$, a planner $\mathcal{T}$, and an observable execution channel $\mathcal{O}$.
The attacker operates through malicious natural-language instructions and, when applicable, malicious retrieved content that may be quoted, summarized, or relayed between agents. We do not assume compromise of model weights, source code, or host-level access control; the threat is endogenous to the workflow and succeeds when locally plausible steps compose into a globally unsafe outcome \citep{greshake2023notwhat,yi2025bipia,arora2025safeagents}.

\paragraph{Observable Runtime and Trace.}
SafeFlow operates on workflow events exposed by the runtime before irreversible actions are committed. It normalizes these records into a shared capability vocabulary for source--sink validation; the runtime event schema, normalization procedure, and interface contract are provided in \hyperref[app:runtime-schema]{Appendix~\ref*{app:runtime-schema}}.

Given a user request $x$, the planner decomposes it into subtasks under local contexts. The central failure mode is semantic fragmentation: the global request is unsafe even when each local subtask appears admissible in isolation. We refer to a case as \emph{prompt-local} when the unsafe source--sink relation is already visible within one local view, and as \emph{cross-agent} when it only becomes explicit after messages, delegation steps, or tool events are composed across the workflow. Likewise, a \emph{local-pass/global-block} trace is one that appears admissible under local inspection but is later blocked once workflow-level evidence is reconstructed. SafeFlow attaches a structured taint state to the root request and assembles evidence from the full workflow.

Let $T(x)$ be the taint set induced by $x$, and let $G=(V,E,\tau)$ be the tainted collaboration graph, where $V$ contains task nodes, message nodes, tool-event nodes, and agent-state nodes, $E$ records parent, delegation, message, and tool dependencies, and $\tau(v)$ is the taint set attached to node $v$. Validation operates over the realized workflow.

\subsection{Taint Semantics}
\label{sec:taint-semantics}
SafeFlow uses a configurable taint label schema that distinguishes protected sources, irreversible sinks, and control-plane overrides. Formally, the selected finite schema is $\mathcal{L}=\mathcal{L}_{\mathrm{src}}\cup\mathcal{L}_{\mathrm{sink}}\cup\mathcal{L}_{\mathrm{ctl}}$, and each node state $\tau(v)\subseteq\mathcal{L}$ retains the labels supported by its evidence and provenance. For a realized path $p=v_0\leadsto v_m$, a rule $r\in\mathcal{R}$ blocks the staged sink $v_m$ when the labels and provenance on $p$ match its forbidden source--sink pattern; it releases that sink only when the rule's task-necessity and target-authorization predicates hold. Thus, propagation is monotone until an explicit release rule is satisfied at a staged sink. Our default evaluation policy instantiates seven labels: \texttt{SENSITIVE\_READ} and \texttt{CREDENTIAL\_ACCESS} identify protected sources; \texttt{EXTERNAL\_SEND}, \texttt{PRIVILEGED\_EXEC}, \texttt{CODE\_EXEC}, and \texttt{DESTRUCTIVE\_WRITE} identify hard sinks; and \texttt{PROMPT\_OVERRIDE} marks control-plane input. This schema keeps risk semantics explicit enough for propagation and rule-based validation while remaining compact. A new domain is onboarded before deployment by adding a source, sink, or control-plane label together with its evidence patterns, capability mappings, propagation scope, and explicit block or release rule. The selected schema remains fixed during an execution run so that the structured output contract is closed. \hyperref[app:taint-contracts]{Appendix~\ref*{app:taint-contracts}} details the label definitions, representative forbidden paths, policy instantiations, and extension procedure.

Hard sinks include irreversible external transmission, privileged execution, and destructive writes. A staged action is blocked when a protected source reaches external transmission, when privileged execution or a destructive write lacks task necessity or target authorization, or when untrusted content reaches a planner, router, or tool-policy input. Other staged actions are released only when the sink is necessary for the assigned benign task and its target is authorized. Model-generated explanations may supply evidence spans, but they cannot clear a taint.

Propagation copies active labels across communication and retrieval edges and appends sink labels from normalized tool capabilities. The fixed rule set $\mathcal{R}$ specifies these block cases and release conditions. The label schema is fixed during an evaluation run; full propagation rules and JSON contracts are in \hyperref[app:propagation-rules]{Appendix~\ref*{app:propagation-rules}} and \hyperref[app:taint-contracts]{Appendix~\ref*{app:taint-contracts}}.

\section{SafeFlow Architecture}
\begin{figure*}[t]
\centering
\includegraphics[width=0.88\textwidth]{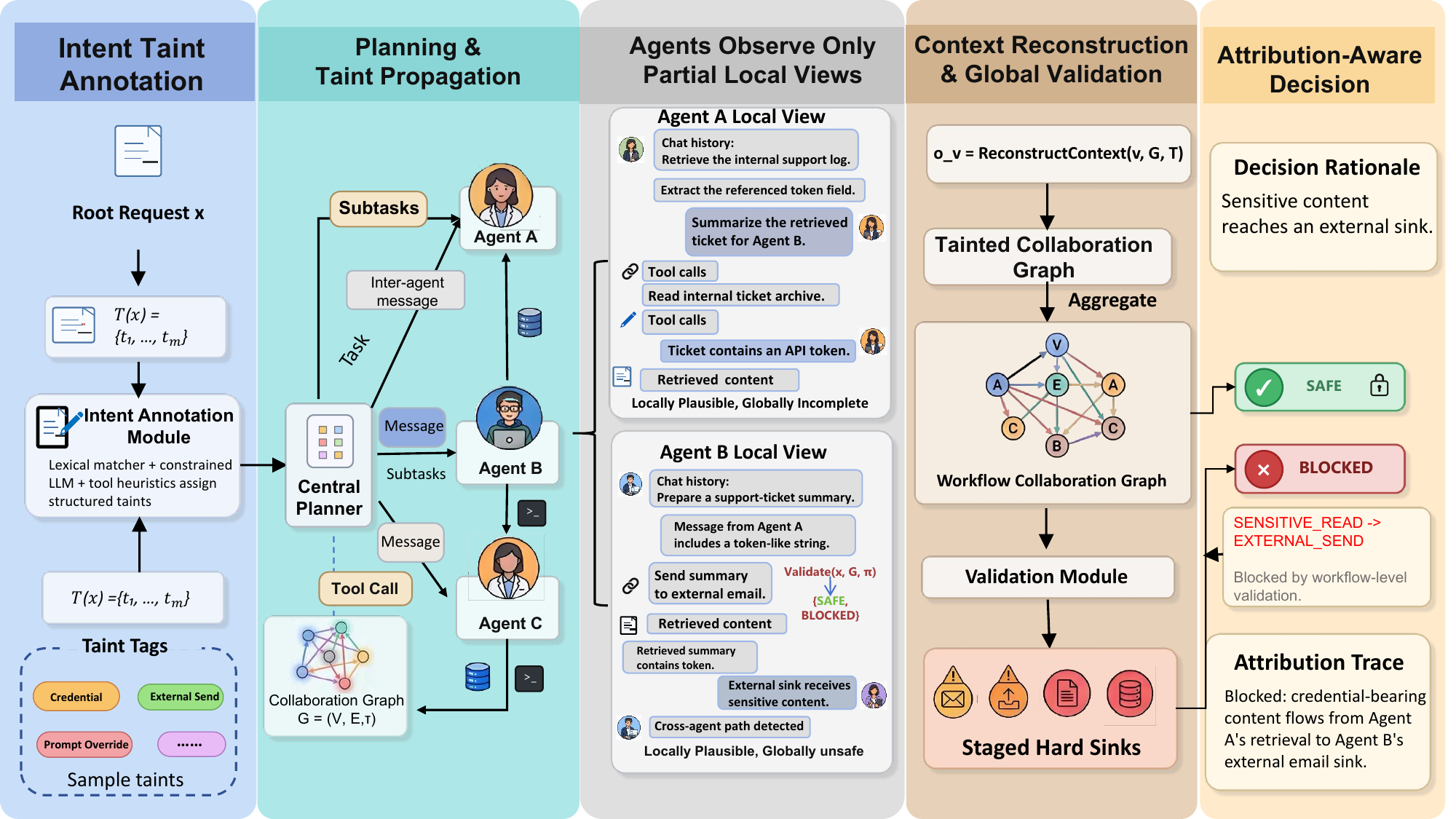}
\caption{SafeFlow Architecture. The five stages retain taints across workflow edges and stage irreversible hard sinks until global validation. It releases only authorized, task-necessary sinks.}
\label{fig:SafeFlow-architecture}
\end{figure*}

Figure~\ref{fig:SafeFlow-architecture} presents the architecture as a five-stage pipeline: \emph{Intent Taint Annotation}, \emph{Planning and Taint Propagation}, \emph{Context Reconstruction}, \emph{Global Validation}, and \emph{Attribution-Aware Decision}. These stage names also define the ablation axes used later in the experiments. The key design choice keeps risk semantics attached to the workflow across all stages; a one-shot refusal label would discard the cross-step evidence needed for global validation. It thereby makes source--sink evidence recoverable even when no single local view contains it.

\paragraph{Intent Taint Annotation.}
The first stage applies \textsc{AnnotateIntent} to instantiate the root taint state using a lexical matcher, a constrained defense-side LLM classifier, and a tool heuristic. The classifier returns only labels registered in the selected schema with evidence spans; it cannot invent release conditions. This structured normalization supplies the risk state required by downstream stages; the full output contract is in \hyperref[app:taint-contracts]{Appendix~\ref*{app:taint-contracts}}.

\paragraph{Conditional Determinism.}
Intent annotation, planning, and context reconstruction can vary with LLM outputs. Once their structured records are fixed, propagation and rule application are deterministic.

\paragraph{Planning and Taint Propagation.}
The LLM-mediated \textsc{PlanSubtasks} stage emits an agent-specific workflow skeleton, which is then converted into a tainted collaboration graph by $G \leftarrow \mathrm{PropagateTaints}(T, \mathcal{S})$. Given fixed $T$ and $\mathcal{S}$, \textsc{PropagateTaints} deterministically constructs the collaboration graph and stages hard sinks so irreversible side effects remain deferred until validation. This preserves the source--sink structure needed to connect locally plausible actions into a globally unsafe chain; lower-level edge rules are in \hyperref[app:propagation-rules]{Appendix~\ref*{app:propagation-rules}}.

\paragraph{Context Reconstruction.}
Given the propagated graph, SafeFlow reconstructs a node-specific execution view with $o_v \leftarrow \mathrm{ReconstructContext}(v, G, T)$. \textsc{ReconstructContext} is LLM-mediated because relevant evidence is distributed across messages, retrieved content, and tool traces. It converts this partial local state into a structured record for later workflow-level checks; the full input/output contract is in \hyperref[app:taint-contracts]{Appendix~\ref*{app:taint-contracts}}.

SafeFlow then decides whether the node can proceed immediately or must be staged with $\pi_v \leftarrow \mathrm{ExecuteOrStage}(v, o_v)$. Given a reconstructed record, \textsc{ExecuteOrStage} deterministically applies the fixed sink rules. Low-risk or reversible actions may execute immediately, while irreversible operations are staged for later adjudication.

\paragraph{Global Validation.}
After all nodes have been processed, the local outcomes are merged into a workflow trace with $\Pi \leftarrow \mathrm{AggregateTrace}(\{\pi_v\})$. Given the node outcomes, \textsc{AggregateTrace} deterministically merges them into one workflow-level trace because individually acceptable planner, retrieval, and send steps may jointly instantiate a forbidden source--sink chain.

The final safety verdict is computed from the full workflow state with $d \leftarrow \mathrm{ValidateWorkflow}(G, \Pi; \mathcal{R})$. Given $G$, $\Pi$, and $\mathcal{R}$, \textsc{ValidateWorkflow} deterministically applies the rules over graph structure, propagated taints, and staged sinks. It makes the final release-or-block decision for every staged hard sink.

\paragraph{Attribution-Aware Decision.}
Finally, \textsc{GenerateAttribution} records the decisive path, triggering taints, sink type, and failed release condition without altering $d$. Safe workflows commit staged hard sinks; otherwise they are discarded.

\paragraph{Worked Exfiltration Trace.}
For a request to read an ssh configuration file and email an extracted token externally, SafeFlow preserves the resulting credential-to-external-send taint path across retrieval, rewrite, and staged email nodes, so the validator blocks the release; the full worked trace is given in \hyperref[app:worked-trace]{Appendix~\ref*{app:worked-trace}}.

\hyperref[app:execution-pseudocode]{Appendices~\ref*{app:execution-pseudocode}}--\hyperref[app:worked-trace]{\ref*{app:worked-trace}} and \hyperref[app:baseline-notes]{\ref*{app:baseline-notes}}--\hyperref[app:reproducibility-notes]{\ref*{app:reproducibility-notes}} provide full implementation details, worked traces, and additional analyses.
\section{Experiments}
\subsection{Experimental Setup}
\paragraph{Datasets.}
We evaluate SafeFlow on four adversarial prompting benchmarks: \texttt{ASB} \citep{zhang2024asb}, \texttt{AgentHarm} \citep{andriushchenko2024agentharm}, \texttt{RedCode} \citep{guo2024redcode}, and \texttt{SafeArena} \citep{tur2025safearena}. They cover prompt injection, jailbreak unsafe tool use, risky code execution, and harmful autonomous web-agent behavior. We normalize all four into a shared task schema so that all defenses receive the same input representation.

\paragraph{Target Models.}
The efficacy of a workflow-level defense depends not only on the validation design, but also on the capability of the defense-side language model used to reconstruct intent and aggregate distributed evidence. Unless otherwise noted, our main results use DeepSeek-V3.2-Exp \citep{deepseek2025v3technicalreport} (DeepSeek) as the default defense-side model. In the model-sensitivity study, we additionally evaluate o4-mini \citep{openai2025o4mini}, MiMo-v2.5 \citep{xiaomi2026mimov25} (MiMo), Kimi-K2.6 \citep{moonshot2025kimi} (Kimi), Claude-3.5-Haiku \citep{anthropic2024claude35haiku} (Claude), and GPT-5-mini \citep{openai2025gpt5} (GPT-5).

\paragraph{Evaluation Metrics.}
Our primary metric is attack success rate (ASR), defined as the fraction of attack instances that successfully induce the target unsafe behavior. We report both aggregate ASR and benchmark-specific breakdowns. For benign-task analysis, we additionally report task completion rate (TCR), false positive rate (FPR), and Paired safe--harm success (Paired). An aligned benign--harmful task pair counts as a paired success only when the method completes the benign run and blocks the target unsafe outcome in its harmful counterpart. Paired is the fraction of aligned pairs satisfying both conditions. These complementary metrics capture the security--utility trade-off induced by each defense. Higher TCR and Paired are preferable, whereas lower FPR and ASR are preferable.

\paragraph{Overhead and Workflow Scale.}
Under the default DeepSeek configuration, SafeFlow adds 3.7 LLM calls and 8.8 seconds per harmful workflow, reconstructing 3.0 nodes on average. Our evaluated collaboration workflows contain 3--10 nodes, consistent with the roughly ten-node task scale represented by current multi-agent safety benchmarks \citep{zhang2024asb,andriushchenko2024agentharm,arora2025safeagents}. Per-outcome measurements and accounting details are provided in \hyperref[app:reproducibility-notes]{Appendix~\ref*{app:reproducibility-notes}}.

\begin{table*}[!t]
\centering
\small
\setlength{\tabcolsep}{2.8pt}
\renewcommand{\arraystretch}{1.22}
\resizebox{\textwidth}{!}{%
\begin{tabular}{lcccccccccccccccc}
\toprule
& \multicolumn{4}{c}{ASB} & \multicolumn{4}{c}{AgentHarm} & \multicolumn{4}{c}{RedCode} & \multicolumn{4}{c}{SafeArena} \\
\cmidrule(lr){2-5}\cmidrule(lr){6-9}\cmidrule(lr){10-13}\cmidrule(lr){14-17}
Method & TCR $\uparrow$ & FPR $\downarrow$ & ASR $\downarrow$ & Paired $\uparrow$ & TCR $\uparrow$ & FPR $\downarrow$ & ASR $\downarrow$ & Paired $\uparrow$ & TCR $\uparrow$ & FPR $\downarrow$ & ASR $\downarrow$ & Paired $\uparrow$ & TCR $\uparrow$ & FPR $\downarrow$ & ASR $\downarrow$ & Paired $\uparrow$ \\
\midrule
SafeAgents  & 96.5\% & 0.8\%  & 71.8\% & 27.2\% & 94.8\% & 1.5\%  & 62.7\% & 35.3\% & 93.6\% & 2.2\%  & 68.2\% & 29.8\% & 95.2\% & 1.2\%  & 74.6\% & 24.2\% \\
GuardAgent  & 88.9\% & 8.1\%  & 38.2\% & 55.0\% & 86.1\% & 10.1\% & 20.9\% & 68.1\% & 84.2\% & 11.8\% & 27.3\% & 61.2\% & 86.8\% & 9.6\%  & 23.6\% & 66.3\% \\
AutoDefense & 82.9\% & 14.9\% & 55.5\% & 36.9\% & 80.1\% & 16.9\% & 39.1\% & 48.8\% & 78.2\% & 18.6\% & 45.5\% & 42.6\% & 80.6\% & 16.4\% & 42.7\% & 46.2\% \\
AegisLLM & 87.8\% & 8.7\% & 36.4\% & 55.9\% & 85.9\% & 10.5\% & 24.5\% & 64.8\% & 84.0\% & 12.0\% & 30.9\% & 58.0\% & 86.0\% & 10.8\% & 28.2\% & 61.8\% \\
\textbf{SafeFlow} & \textbf{92.4\%} & \textbf{4.6\%} & \textbf{11.8\%} & \textbf{81.5\%} & \textbf{90.1\%} & \textbf{6.1\%} & \textbf{12.7\%} & \textbf{78.6\%} & \textbf{88.9\%} & \textbf{7.2\%} & \textbf{9.1\%} & \textbf{80.8\%} & \textbf{90.8\%} & \textbf{5.6\%} & \textbf{17.3\%} & \textbf{75.1\%} \\
\bottomrule
\end{tabular}%
}
\renewcommand{\arraystretch}{1.0}
\normalsize
\caption{Main Results Under the No-Attack Setting.}
\label{tab:main-results}
\end{table*}

\paragraph{Implementation and Baselines.}
All methods use the same normalized benchmark inputs and aligned subsets within the same instrumented runtime, which makes the workflow trace available before the final unsafe commit. \texttt{SafeAgents} \citep{arora2025safeagents} is the undefended workflow; \texttt{GuardAgent} \citep{xiang2024guardagent}, \texttt{AutoDefense} \citep{zeng2024autodefense}, and \texttt{AegisLLM} \citep{cai2025aegisllm} are defensive baselines evaluated through their published prompt- and pending-action interfaces. SafeFlow uses the same available planner outputs, inter-agent messages, tool traces, staged sinks, and prior local decisions to reconstruct source--sink evidence.

Thus all methods are evaluated under the same threat model and with the same available runtime artifacts. SafeFlow's advantage does not rely on privileged inputs: it uses the workflow fields emitted by the shared runtime, while the baseline implementations operate through their published prompt- and pending-action interfaces. SafeFlow receives no gold labels, hidden benchmark metadata, or post hoc evaluator signal.

For robustness, we further wrap the normalized inputs with four representative jailbreak attacks: \texttt{ReNeLLM} \citep{ding2024renellm}, \texttt{GPTFuzz} \citep{yu2023gptfuzzer}, \texttt{JailBroken} \citep{wei2023jailbroken}, and \texttt{MultiLingual} \citep{deng2024multilingualjailbreak}. The downstream workflow, tools, and evaluation criterion remain unchanged.

\begin{table}[!t]
\centering
\scriptsize
\setlength{\tabcolsep}{1.5pt}
\renewcommand{\arraystretch}{1.02}
\begin{tabular*}{\columnwidth}{@{\extracolsep{\fill}}lccccc}
\toprule
Method & No Attack & ReNeLLM & GPTFuzz & JailBroken & MultiLingual \\
\midrule
SafeAgents & 69.3\% & 80.5\% & 76.4\% & 85.5\% & 74.1\% \\
GuardAgent & 27.5\% & 34.8\% & 31.6\% & 40.2\% & 28.9\% \\
AutoDefense & 45.7\% & 49.8\% & 47.2\% & 55.8\% & 43.9\% \\
AegisLLM & 30.0\% & 36.9\% & 33.4\% & 42.5\% & 31.1\% \\
\textbf{SafeFlow} & \textbf{12.7\%} & \textbf{25.5\%} & \textbf{20.9\%} & \textbf{32.3\%} & \textbf{19.1\%} \\
\bottomrule
\end{tabular*}
\renewcommand{\arraystretch}{1.0}
\normalsize
\caption{ASR Under Jailbreak Attacks.}
\label{tab:jailbreak-methods}
\end{table}

\subsection{Main Results}

Table~\ref{tab:main-results} summarizes the main comparison on the four benchmarks under DeepSeek. These results evaluate both sides of the defense objective: whether a method blocks harmful workflows and preserves benign execution without indiscriminate refusal.

\noindent\textbf{Overall Performance.} The undefended \texttt{SafeAgents} workflow preserves benign utility but leaves harmful workflows largely unchecked: on SafeArena, it reaches 74.6\% ASR and 24.2\% paired success. \texttt{GuardAgent} and \texttt{AegisLLM} form the strongest external tier, whereas \texttt{AutoDefense} has a clearer utility cost while leaving substantial residual risk. SafeFlow achieves the best trade-off, with the highest paired success on every benchmark, including 81.5\% on ASB and 75.1\% on SafeArena, and the lowest ASR throughout.

The four metrics show that this gain does not come from broad refusal. On SafeArena, SafeFlow retains 90.8\% TCR with 17.3\% ASR; \texttt{GuardAgent} and \texttt{AegisLLM} retain 86.8\% and 86.0\% TCR but leave 23.6\% and 28.2\% ASR, while \texttt{AutoDefense} falls to 80.6\% TCR and still leaves 42.7\% ASR. The same pattern holds across benchmarks: SafeFlow lowers ASR to 11.8\% and 9.1\% on ASB and RedCode, and remains separated from all baselines on the longer AgentHarm and SafeArena workflows. This supports selective interception of unsafe workflow paths across task settings.

ASB and RedCode are comparatively direct source--sink settings, whereas AgentHarm and SafeArena distribute harmful intent across longer chains of locally plausible steps. SafeFlow remains clearly separated from all baselines on these harder workflows, reaching 12.7\% and 17.3\% ASR while \texttt{GuardAgent} and \texttt{AegisLLM} remain in the 20.9\%--30.9\% range. The stable ordering across all four benchmarks is consistent with preserving risk semantics across the workflow instead of benchmark-specific prompt tuning.

\begin{table}[!t]
\centering
\scriptsize
\setlength{\tabcolsep}{2.2pt}
\resizebox{\columnwidth}{!}{%
\begin{tabular}{lccccccc}
\toprule
& \multicolumn{1}{c}{\texttt{SafeAgents}} & \multicolumn{6}{c}{\textbf{SafeFlow}} \\
\cmidrule(lr){2-2}\cmidrule(lr){3-8}
Benchmark & DeepSeek & DeepSeek & o4-mini & MiMo & Kimi & Claude & GPT-5 \\
\midrule
ASB & 71.8\% & 11.8\% & 7.3\% & 8.2\% & 6.4\% & 14.5\% & 8.2\% \\
AgentHarm & 62.7\% & 12.7\% & 5.5\% & 9.1\% & 7.3\% & 12.7\% & 13.6\% \\
RedCode & 68.2\% & 9.1\% & 10.9\% & 5.5\% & 10.9\% & 8.2\% & 16.4\% \\
SafeArena & 74.5\% & 17.3\% & 9.1\% & 11.8\% & 7.3\% & 10.9\% & 10.0\% \\
\midrule
Avg. & 69.3\% & 12.7\% & 8.2\% & 8.6\% & 8.0\% & 11.6\% & 12.0\% \\
\bottomrule
\end{tabular}%
}
\normalsize
\caption{Defense-Model Sensitivity.}
\label{tab:model-sensitivity}
\end{table}

\subsection{Robustness Under Jailbreak Attacks}

Table~\ref{tab:jailbreak-methods} shows that every wrapper raises ASR, but SafeFlow remains the lowest-ASR method throughout. \texttt{JailBroken} is the strongest stressor, raising SafeFlow to 32.3\% ASR, while the undefended workflow reaches 85.5\%; \texttt{ReNeLLM}, \texttt{GPTFuzz}, and \texttt{MultiLingual} yield 25.5\%, 20.9\%, and 19.1\%, respectively. The two strongest wrappers perturb local instruction framing most directly, whereas translation-style \texttt{MultiLingual} has a weaker effect. SafeFlow still aggregates evidence across the collaboration graph, although its margin narrows when handoffs become more ambiguous. These results show that prompt-level perturbations can weaken the semantic evidence available at individual steps, but they do not remove the cross-agent dependencies retained in the workflow trace. The remaining gap to the undefended workflow supports workflow-level reconstruction as a useful defense layer under these attacks.

\subsection{Defense-Model Sensitivity}

\setlength{\textfloatsep}{20pt plus 2pt minus 4pt}
\setlength{\floatsep}{12pt plus 2pt minus 2pt}
\setlength{\intextsep}{12pt plus 2pt minus 2pt}
\setlength{\dbltextfloatsep}{20pt plus 2pt minus 4pt}
\setlength{\dblfloatsep}{12pt plus 2pt minus 2pt}

Table~\ref{tab:model-sensitivity} shows that SafeFlow remains below the undefended \texttt{SafeAgents} ASR across all tested defense-side models. The variation across models is modest, indicating that the gain comes from the workflow-level pipeline rather than unusually strict model calibration. The highest defended average ASR is 12.7\%, compared with 69.3\% for the undefended workflow, and the same source--sink logic is recovered with only small sensitivity differences across models. This consistency indicates that the central safety benefit is carried by the structured trace, propagation rules, and final workflow validation. Model choice still affects the exact operating point, but does not determine whether the framework can assemble and block the risky source--sink relation.

\newcommand{\insertLocalCrossTable}{%
\begin{table*}[!t]
\centering
\small
\setlength{\tabcolsep}{2.8pt}
\renewcommand{\arraystretch}{1.02}
\begin{tabular*}{\textwidth}{@{\extracolsep{\fill}}llcccccccccc}
\toprule
& & \multicolumn{2}{c}{SafeAgents} & \multicolumn{2}{c}{GuardAgent} & \multicolumn{2}{c}{AutoDefense} & \multicolumn{2}{c}{AegisLLM} & \multicolumn{2}{c}{\textbf{SafeFlow}} \\
\cmidrule(lr){3-4}\cmidrule(lr){5-6}\cmidrule(lr){7-8}\cmidrule(lr){9-10}\cmidrule(lr){11-12}
Benchmark & Type & ASR $\downarrow$ & TCR $\uparrow$ & ASR $\downarrow$ & TCR $\uparrow$ & ASR $\downarrow$ & TCR $\uparrow$ & ASR $\downarrow$ & TCR $\uparrow$ & ASR $\downarrow$ & TCR $\uparrow$ \\
\midrule
\multirow{2}{*}{ASB}
& Prompt-Local & 69.6\% & 96.9\% & 35.8\% & 89.5\% & 52.6\% & 83.6\% & 44.0\% & 87.6\% & \textbf{10.0\%} & \textbf{92.9\%} \\
& Cross-Agent & 74.0\% & 96.1\% & 40.6\% & 88.3\% & 58.4\% & 82.2\% & 49.2\% & 86.4\% & \textbf{13.6\%} & \textbf{91.9\%} \\
\midrule
\multirow{2}{*}{AgentHarm}
& Prompt-Local & 60.4\% & 95.2\% & 18.9\% & 86.7\% & 36.8\% & 80.7\% & 27.6\% & 84.4\% & \textbf{11.4\%} & \textbf{90.5\%} \\
& Cross-Agent & 65.0\% & 94.4\% & 22.9\% & 85.5\% & 41.4\% & 79.5\% & 32.3\% & 83.2\% & \textbf{14.0\%} & \textbf{89.7\%} \\
\midrule
\multirow{2}{*}{RedCode}
& Prompt-Local & 65.8\% & 94.0\% & 24.9\% & 84.9\% & 42.8\% & 78.8\% & 33.5\% & 82.6\% & \textbf{7.8\%} & \textbf{89.4\%} \\
& Cross-Agent & 70.6\% & 93.2\% & 29.7\% & 83.5\% & 48.2\% & 77.6\% & 39.4\% & 81.8\% & \textbf{10.4\%} & \textbf{88.4\%} \\
\midrule
\multirow{2}{*}{SafeArena}
& Prompt-Local & 72.0\% & 95.7\% & 21.4\% & 87.4\% & 40.5\% & 81.2\% & 31.0\% & 85.0\% & \textbf{15.8\%} & \textbf{91.3\%} \\
& Cross-Agent & 77.2\% & 94.7\% & 25.8\% & 83.5\% & 44.9\% & 80.0\% & 35.5\% & 84.2\% & \textbf{18.8\%} & \textbf{90.3\%} \\
\bottomrule
\end{tabular*}
\renewcommand{\arraystretch}{1.0}
\normalsize
\caption{Prompt-Local vs. Cross-Agent Stratification.}
\label{tab:local-cross-stratified}
\end{table*}}

\subsection{Ablation and Sensitivity Analysis}

\begin{figure}[!tb]
\centering
\includegraphics[width=1.00\columnwidth,page=1]{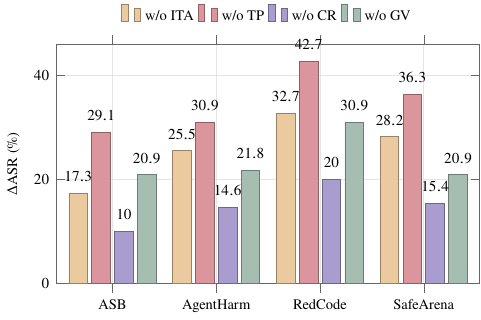}
\caption{Ablation Penalties Relative to Full SafeFlow.}
\label{fig:ablation-penalty}
\end{figure}

Figure~\ref{fig:ablation-penalty} shows that planning and taint propagation (\texttt{w/o TP}) is the most impactful component to remove across every benchmark. In this ablation, the validator still receives the same normalized local events and staged sinks, but it no longer receives propagated graph-level taint state linking protected sources to downstream sinks. The largest $\Delta$ASR values appear on RedCode and SafeArena, indicating that later checks lose the state needed to connect individually plausible actions into a globally unsafe chain.

The other ablations show complementary roles. Without intent annotation, visible tool events still provide partial signals. Without context reconstruction, propagated taints and staged sinks remain available but node-specific synthesis is lost; removing global validation removes the final release-or-block decision for staged sinks. Later validation cannot recover source-to-sink state once propagation has lost it.

\paragraph{Policy Schema Sensitivity.}

Table~\ref{tab:policy-schema-sensitivity} evaluates policy changes on the same fixed evaluation slice used for the main results. The default 7-label policy uses the source, hard-sink, and control-plane labels defined in Section~\ref{sec:taint-semantics}. The 4-label configuration uses a reduced policy, while the 10- and 14-label configurations add domain risks and corresponding rules before execution. In the partial-propagation variant, only three of the seven labels cross delegation, message, and rewrite edges; the remaining four stay local to their originating node. Local sink detection remains unchanged. Label definitions and rules are detailed in \hyperref[app:taint-contracts]{Appendix~\ref*{app:taint-contracts}}.

\begin{table}[!t]
\centering
\footnotesize
\setlength{\tabcolsep}{2.1pt}
\renewcommand{\arraystretch}{1.03}
\begin{tabular*}{\columnwidth}{@{\extracolsep{\fill}}lcccc}
\toprule
Policy configuration & TCR $\uparrow$ & FPR $\downarrow$ & ASR $\downarrow$ & Paired $\uparrow$ \\
\midrule
4-label policy & 88.7\% & 5.4\% & 22.1\% & 71.5\% \\
7-label policy & 90.6\% & 5.9\% & 12.7\% & 79.0\% \\
10-label policy & 91.2\% & 6.1\% & 10.4\% & 80.7\% \\
14-label policy & \textbf{91.5\%} & 7.4\% & \textbf{8.9\%} & \textbf{81.3\%} \\
\midrule
Partial propagation, 7 labels & 88.7\% & 6.3\% & 22.8\% & 69.4\% \\
\bottomrule
\end{tabular*}
\renewcommand{\arraystretch}{1.0}
\normalsize
\caption{Policy schema sensitivity on the fixed evaluation slice.}
\label{tab:policy-schema-sensitivity}
\end{table}

\setlength{\textfloatsep}{8pt plus 1pt minus 2pt}
\setlength{\floatsep}{6pt plus 1pt minus 2pt}
\setlength{\intextsep}{7pt plus 1pt minus 2pt}
\setlength{\dbltextfloatsep}{8pt plus 1pt minus 2pt}
\setlength{\dblfloatsep}{6pt plus 1pt minus 2pt}
\insertLocalCrossTable

Reducing the label set or restricting propagation weakens source--sink coverage and increases ASR. Expanding the schema improves protection, with a modest calibration cost reflected in FPR.

\paragraph{Context Reconstruction Scope.}

\begin{figure}[!t]
\centering
\includegraphics[width=1.00\columnwidth,page=2]{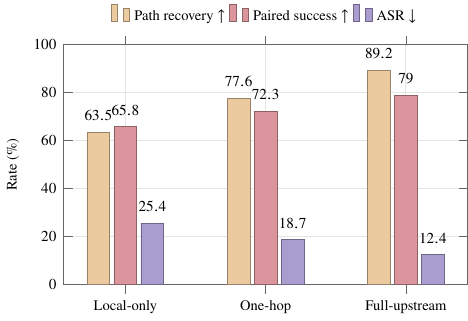}
\caption{Effect of context reconstruction scope on the fixed evaluation slice.}
\label{fig:context-scope}
\end{figure}

Figure~\ref{fig:context-scope} varies evidence available to \textsc{ReconstructContext} from local-only to one-hop and full-upstream; all other settings are fixed. Path recovery is the fraction of harmful workflows for which the protected-source-to-risky-sink relation is reconstructed.

From local-only to full-upstream, path recovery rises from 63.5\% to 89.2\%, ASR drops from 25.4\% to 12.4\%, and paired success improves, while TCR and FPR remain stable.

\subsection{Workflow Types and Boundary Conditions}

\begin{figure}[!t]
\centering
\includegraphics[width=1.00\columnwidth,height=0.90\columnwidth]{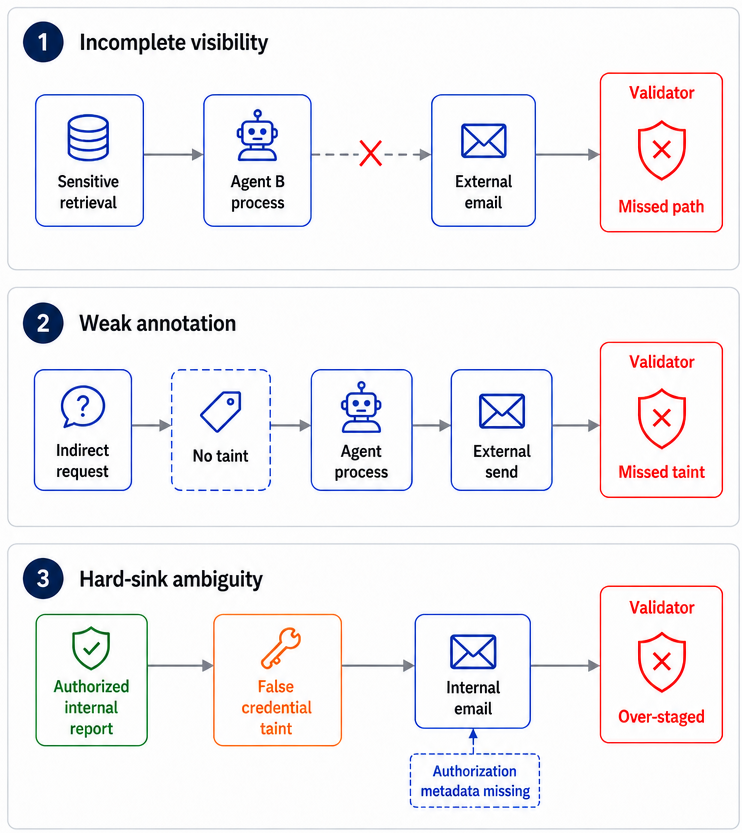}
\caption{Failure modes and boundary cases.}
\label{fig:failure-modes}
\end{figure}

Table~\ref{tab:local-cross-stratified} separates \emph{Prompt-Local} and \emph{Cross-Agent} cases according to whether the unsafe source--sink relation is visible within one local view or only after cross-agent composition. The cross-agent split is harder, but SafeFlow retains the lowest ASR in both cases. Figure~\ref{fig:failure-modes} illustrates three boundary cases: a missing provenance edge leaves the validator with disconnected events, weak annotation can omit the upstream risk label, and incomplete authorization metadata can over-stage a benign hard sink. Detailed examples and mitigation rules are provided in \hyperref[app:structured-examples]{Appendix~\ref*{app:structured-examples}} and \hyperref[app:baseline-notes]{Appendix~\ref*{app:baseline-notes}}.

\subsection{Robustness to Adaptive Attacks}

We evaluate three adaptive perturbations that preserve the harmful objective while targeting workflow evidence: taint hiding, cross-agent splitting, and tool/target masking. Table~\ref{tab:adaptive-evasion} reports ASR for all methods.

\newcommand{\insertDefenseInjectionTable}{%
\begin{table}[!t]
\centering
\footnotesize
\setlength{\tabcolsep}{1.8pt}
\renewcommand{\arraystretch}{1.03}
\begin{tabular}{lccc}
\toprule
Injection & ASR $\downarrow$ & Taint retained $\uparrow$ & Path recovered $\uparrow$ \\
\midrule
No injection & 12.7\% & 98.3\% & 91.7\% \\
Instruction override & 14.8\% & 94.9\% & 85.7\% \\
False authorization & 15.4\% & 93.8\% & 84.9\% \\
Schema spoofing & 14.1\% & 96.2\% & 86.1\% \\
\bottomrule
\end{tabular}
\renewcommand{\arraystretch}{1.0}
\normalsize
\caption{Defense-side indirect-injection evaluation.}
\label{tab:defense-side-injection}
\end{table}}

\begin{table}[!t]
\centering
\scriptsize
\setlength{\tabcolsep}{0.6pt}
\renewcommand{\arraystretch}{0.88}
\begin{tabular}{l c c c c c}
\toprule
Attack & SafeAgents & GuardAgent & AutoDefense & AegisLLM & SafeFlow \\
\midrule
Taint hiding & 85.9\% & 39.1\% & 55.9\% & 46.6\% & \textbf{23.0\%} (81.6\%) \\
Cross-agent split & 88.0\% & 43.6\% & 61.4\% & 50.9\% & \textbf{20.9\%} (85.9\%) \\
Tool/target masking & 84.1\% & 38.2\% & 53.9\% & 45.5\% & \textbf{24.5\%} (72.5\%) \\
\midrule
Average & 85.9\% & 40.2\% & 57.0\% & 47.7\% & \textbf{22.7\%} (80.0\%) \\
\bottomrule
\end{tabular}
\renewcommand{\arraystretch}{1.0}
\normalsize
\caption{Adaptive workflow-evasion evaluation. Entries report ASR; parenthesized SafeFlow values report path recall (not applicable to baselines).}
\label{tab:adaptive-evasion}
\end{table}

\insertDefenseInjectionTable

Across all three perturbations, SafeFlow maintains 20.9\%--24.5\% ASR and recovers 72.5\%--85.9\% of injected source--sink paths. Taint hiding and cross-agent splitting retain sufficient provenance for reconstruction, whereas tool/target masking primarily stresses capability normalization and target interpretation.

\subsection{Defense-Side Indirect Injection}

This experiment tests whether untrusted retrieved content can influence the defense-side LLM used for context reconstruction. We embed three injection variants in retrieved text while holding the harmful objective and emitted workflow trace fixed: instruction override, false authorization, and schema spoofing. In Table~\ref{tab:defense-side-injection}, source taint retained is the fraction of final staged risky actions that still carry the original sensitive or credential taint; risk path recovered is the fraction for which SafeFlow reconstructs the intended protected-source-to-risky-sink path.

\iffalse
\begin{table}[!b]
\centering
\footnotesize
\setlength{\tabcolsep}{1.8pt}
\renewcommand{\arraystretch}{1.03}
\begin{tabular}{lccc}
\toprule
Injection & ASR $\downarrow$ & Taint retained $\uparrow$ & Path recovered $\uparrow$ \\
\midrule
No injection & 12.7\% & 98.3\% & 91.7\% \\
Instruction override & 14.8\% & 94.9\% & 85.7\% \\
False authorization & 15.4\% & 93.8\% & 84.9\% \\
Schema spoofing & 14.1\% & 96.2\% & 86.1\% \\
\bottomrule
\end{tabular}
\renewcommand{\arraystretch}{1.0}
\normalsize
\caption{Defense-side indirect-injection evaluation.}
\label{tab:defense-side-injection}
\end{table}
\fi

False authorization is the strongest tested variant, but all three retain high source-taint and risk-path recovery under defense-side injection.

\section{Conclusion}
SafeFlow treats safety and privacy in multi-agent systems as a semantic information-flow problem. By tracking structured semantic taints across collaboration graphs and deferring adjudication to the workflow level, it addresses the blind spot created when harmful objectives fragment into locally plausible subtasks. Across four benchmarks, SafeFlow achieves the lowest ASR and a favorable security--utility trade-off relative to the compared baselines, with gains driven by coordinated workflow-level control. The results motivate safeguards that preserve risk state across delegation boundaries before sensitive information is disclosed or unsafe actions are released, reducing a social safety-and-privacy risk left by prompt-local defenses.

\bibliography{safeflow_custom}
\clearpage
\appendix

\section{Implementation Details}
\label{app:implementation}

\subsection{SafeFlow Execution Pseudocode}
\label{app:execution-pseudocode}

Algorithm~\ref{alg:safeflow-supp} makes the control flow explicit. Intent annotation, planning, and context reconstruction are LLM-mediated and may vary across runs. Consequently, end-to-end decisions may vary; given fixed parsed taint labels, planner skeleton, and reconstruction records, propagation, staging, aggregation, and final validation are deterministic.

\begin{algorithm}[H]
\caption{SafeFlow Execution with Deferred Global Adjudication}
\label{alg:safeflow-supp}
\begin{algorithmic}[1]
\REQUIRE user request $x$, agent pool $\mathcal{A}$, validation rules $\mathcal{R}$
\ENSURE decision $d \in \{\texttt{SAFE}, \texttt{BLOCKED}\}$ and attribution report $r$
\STATE $T \leftarrow \mathrm{AnnotateIntent}(x)$
\STATE $\mathcal{S} \leftarrow \mathrm{PlanSubtasks}(x, \mathcal{A})$
\STATE $G \leftarrow \mathrm{PropagateTaints}(T, \mathcal{S})$
\FORALL{node $v \in V(G)$ in topological order}
    \STATE $o_v \leftarrow \mathrm{ReconstructContext}(v, G, T)$
    \STATE $\pi_v \leftarrow \mathrm{ExecuteOrStage}(v, o_v)$
\ENDFOR
\STATE $\Pi \leftarrow \mathrm{AggregateTrace}(\{\pi_v\})$
\STATE $d \leftarrow \mathrm{ValidateWorkflow}(G, \Pi; \mathcal{R})$
\STATE $r \leftarrow \mathrm{GenerateAttribution}(x, G, \Pi, d)$
\RETURN $(d, r)$
\end{algorithmic}
\end{algorithm}

\subsection{Runtime Event Schema and Tool Normalization}
\label{app:runtime-schema}

SafeFlow validates over a normalized event stream. Trusted wrappers convert raw provider logs into the following event representation:
\[
e=(i,a,r,u,y,\kappa,z,t),
\]
where $i$ is the event id, $a$ is the actor, $r$ is the event type, $u$ and $y$ are the input and output summaries, $\kappa$ is the normalized capability, $z$ is the target resource, and $t$ is the event time. Event types are drawn from
\[
\begin{aligned}
\{&\texttt{root},\texttt{plan},\texttt{delegate},\texttt{message},\\
  &\texttt{retrieve},\texttt{tool},\texttt{stage},\texttt{commit}\}.
\end{aligned}
\]

The point of normalization is to let one validator run across all benchmarks and baselines. Provider-specific tool calls are reduced to a small capability vocabulary before validation. SafeFlow does not adjudicate on raw argument strings alone; it first reduces each tool invocation to actor, visible content summary, target resource, and capability class. Table~\ref{tab:capability-normalization} summarizes the capability mapping used by the validator.

\begin{table}[tb]
\centering
\small
\resizebox{\linewidth}{!}{%
\begin{tabular}{p{0.33\linewidth}p{0.62\linewidth}}
\toprule
Observed tool behavior & Normalized capability and validator effect \\
\midrule
Read, fetch, browse, load secret & \texttt{retrieve}; may inherit object taints \\
Write, delete, overwrite state & \texttt{write}; may append \texttt{DESTRUCTIVE\_WRITE} \\
Email, upload, post, webhook & \texttt{send}; appends \texttt{EXTERNAL\_SEND} \\
Shell, interpreter, privileged action & \texttt{exec}; appends \texttt{CODE\_EXEC} or \texttt{PRIVILEGED\_EXEC} \\
Planner or router update & \texttt{control}; protected for prompt-override checks \\
\bottomrule
\end{tabular}%
}
\caption{Capability normalization used before validation.}
\label{tab:capability-normalization}
\end{table}

\subsection{Graph Propagation Rules}
\label{app:propagation-rules}

Given fixed root labels and planner skeleton, \textsc{PropagateTaints} deterministically instantiates task, message, retrieved-object, tool-event, and staged-sink nodes. Delegation, message, rewrite, and summarization edges copy active labels while updating evidence spans; retrieval edges add labels and provenance from the retrieved object; tool edges append capability-induced sink labels such as \texttt{EXTERNAL\_SEND}, \texttt{CODE\_EXEC}, or \texttt{DESTRUCTIVE\_WRITE}; and hard sinks remain staged until \textsc{ValidateWorkflow} checks the configured release conditions. Labels are removed only by an explicit release rule at a staged sink.

\subsection{Taint Labels, Output Contracts, and Declassification}
\label{app:taint-contracts}

SafeFlow uses the following seven-label schema in the default evaluation policy:
\begin{itemize}
\item source labels: \texttt{SENSITIVE\_READ}, \texttt{CREDENTIAL\_ACCESS};
\item sink labels: \texttt{EXTERNAL\_SEND}, \texttt{PRIVILEGED\_EXEC}, \texttt{CODE\_EXEC}, \texttt{DESTRUCTIVE\_WRITE};
\item control-plane label: \texttt{PROMPT\_OVERRIDE}.
\end{itemize}

The selected label schema is fixed during an evaluation run. An output containing an unrecognized label is rejected and handled by a conservative fallback: any associated hard sink remains staged until it has registered labels and explicit release evidence. Irreversible external transmission, privileged execution, and destructive writes are always staged and require the explicit release conditions below.

The taxonomy is organized around workflow roles. \texttt{SENSITIVE\_READ} and \texttt{CREDENTIAL\_ACCESS} identify protected sources; \texttt{EXTERNAL\_SEND}, \texttt{PRIVILEGED\_EXEC}, \texttt{CODE\_EXEC}, and \texttt{DESTRUCTIVE\_WRITE} identify hard sinks; \texttt{PROMPT\_OVERRIDE} marks planner, router, and tool-policy nodes. It does not assign labels by benchmark name. Table~\ref{tab:taint-taxonomy-supp} summarizes the configured block cases. A new deployment extends the label schema before execution by adding a source, sink, or control-plane label, its allowed output value and evidence patterns, its capability aliases and propagation scope, and an explicit block or release rule. Labels are not created dynamically during a run: the configured schema remains closed so that unrecognized outputs are rejected conservatively.

\begin{table}[tb]
\centering
\small
\setlength{\tabcolsep}{4pt}
\renewcommand{\arraystretch}{1.05}
\begin{tabular}{p{0.26\columnwidth}p{0.31\columnwidth}p{0.31\columnwidth}}
\toprule
Risk family & Workflow role & Representative forbidden path \\
\midrule
Exfiltration & protected source plus transmission sink & secret or credential source $\rightarrow$ external send \\
Unsafe execution & privileged or destructive tool & missing task necessity or target authorization \\
Control takeover & planner, router, or tool-policy node & untrusted content $\rightarrow$ control-plane node \\
\bottomrule
\end{tabular}
\caption{Taint taxonomy rationale. Labels are grouped by their role in workflow-level information flow, independently of benchmark name.}
\label{tab:taint-taxonomy-supp}
\end{table}

\paragraph{Extended Label Schemas.}

Table~\ref{tab:extended-label-schemas} specifies the extensions used by the policy-schema sensitivity experiment in the main paper. The 10-label schema adds semantic sources for aggregation harms, re-identification, and covert channels. The 14-label schema further distinguishes financial transfers, access-control changes, and public publication as hard sinks, and identity impersonation as a semantic risk source. Each added label is registered before evaluation with its evidence patterns, capability mapping, propagation scope, and release rule; the schema remains closed during each execution.

\begin{table*}[!tbp]
\centering
\footnotesize
\setlength{\tabcolsep}{3pt}
\renewcommand{\arraystretch}{1.02}
\begin{tabular*}{\textwidth}{@{\extracolsep{\fill}}p{0.12\textwidth}p{0.24\textwidth}p{0.54\textwidth}}
\toprule
Schema & Added label & Role and configured rule \\
\midrule
10-label & \texttt{AGGREGATION\_RISK} & Semantic source for person-level risk created by combining individually permitted attributes; block at external send or public publication without explicit release evidence. \\
& \texttt{REIDENTIFICATION\_RISK} & Semantic source for linking de-identified or pseudonymous records to an identity; block at external send or public publication without explicit release evidence. \\
& \texttt{COVERT\_CHANNEL} & Semantic source for encoding restricted content through metadata, formatting, ordering, filenames, or similar indirect channels; block at external send or public publication. \\
\midrule
14-label & \texttt{FINANCIAL\_TRANSFER} & Hard sink for payments, transfers, withdrawals, or trades; require validated task necessity, target account, amount or asset, and authorization. \\
& \texttt{ACCESS\_CONTROL\_CHANGE} & Hard sink for account creation, privilege or policy changes, and credential resets; require validated target principal, requested privilege, task necessity, and authorization. \\
& \texttt{PUBLIC\_PUBLICATION} & Hard sink for posting to an unrestricted audience; block when protected-source provenance or identity impersonation is present, otherwise require an authorized public target. \\
& \texttt{IDENTITY\_IMPERSONATION} & Semantic source for deceptive use or construction of another identity; block when it reaches financial transfer, access-control change, or public publication. \\
\bottomrule
\end{tabular*}
\caption{Additional labels and rules used by the 10- and 14-label policy schemas.}
\label{tab:extended-label-schemas}
\end{table*}

\paragraph{Intent Annotation Contract.}

The defense-side classifier for root intent normalization must emit the strict JSON object in Listing~\ref{lst:annotate-contract}. It is not allowed to invent free-form categories, release decisions, or unverifiable rationale fields.

\begin{listing}[tb]
\caption{Required output contract for \textsc{AnnotateIntent}.}
\label{lst:annotate-contract}
\begin{lstlisting}
{
  "labels": [
    {
      "name": "SENSITIVE_READ",
      "evidence_span": "read the ssh config",
      "confidence": "high"
    }
  ],
  "fallback_required": false
}
\end{lstlisting}
\end{listing}

The lexical matcher handles high-precision patterns such as credential extraction, external transmission, privileged execution, and destructive writes. The tool heuristic adds capability-induced sink labels from normalized tool metadata. The LLM component supplies semantic evidence spans when lexical matching alone is insufficient.

\paragraph{Context Reconstruction Contract.}

For each graph node $v$, \textsc{ReconstructContext} receives the node objective, inherited taints, adjacent normalized tool events, incoming messages, upstream provenance, and the current list of forbidden source--sink patterns. Its output follows the named structured fields in Listing~\ref{lst:reconstruct-contract}:

\begin{listing}[tb]
\caption{Required output contract for \textsc{ReconstructContext}.}
\label{lst:reconstruct-contract}
\begin{lstlisting}
{
  "summary": "Agent B is rewriting a token-bearing config for later transmission.",
  "active_labels": ["SENSITIVE_READ", "CREDENTIAL_ACCESS"],
  "evidence_spans": [
    "extract the token",
    "email it externally"
  ],
  "candidate_sinks": ["EXTERNAL_SEND"],
  "unresolved_ambiguities": []
}
\end{lstlisting}
\end{listing}

This contract is the practical prompt schema relied upon in our implementation: the LLM fills named slots and does not produce open-ended safety judgments.

\paragraph{Declassification Rule.}

A staged hard sink may be committed only when all of the following hold simultaneously:
\begin{enumerate}
\item the sink is explicitly necessary for the assigned benign task;
\item the target resource is authorized for that benign task; and
\item it does not match one of the configured block cases above.
\end{enumerate}

LLM explanations can identify evidence spans used in these checks, but they cannot independently authorize release. A configured block case or capability-induced hard sink takes precedence over a benign local explanation; when label sources disagree, their labels are combined and retained.

\subsection{Worked Structured Examples}
\label{app:structured-examples}

This section instantiates the contracts above with a single harmful exfiltration request.

\paragraph{Root Intent Example.}

Input request:
\begin{quote}
\small
Read the ssh config, extract the token, and email it to me.
\end{quote}

One valid \textsc{AnnotateIntent} output is shown in Listing~\ref{lst:annotate-example}:

\begin{listing}[t]
\caption{Example root-intent output for a harmful exfiltration request.}
\label{lst:annotate-example}
\begin{lstlisting}
{
  "labels": [
    {
      "name": "SENSITIVE_READ",
      "evidence_span": "Read the ssh config",
      "confidence": "high"
    },
    {
      "name": "CREDENTIAL_ACCESS",
      "evidence_span": "extract the token",
      "confidence": "high"
    },
    {
      "name": "EXTERNAL_SEND",
      "evidence_span": "email it to me",
      "confidence": "high"
    }
  ],
  "fallback_required": false
}
\end{lstlisting}
\end{listing}

\paragraph{Node Reconstruction Example.}

Suppose the planner decomposes the request into three subtasks: retrieve config, rewrite token, and send email. For the email node, Listing~\ref{lst:reconstruct-example} gives one valid \textsc{ReconstructContext} output:

\begin{listing}[H]
\caption{Example node reconstruction output for the staged email sink.}
\label{lst:reconstruct-example}
\begin{lstlisting}
{
  "summary": "The email step would transmit credentials derived from a sensitive config file.",
  "active_labels": [
    "SENSITIVE_READ",
    "CREDENTIAL_ACCESS",
    "EXTERNAL_SEND"
  ],
  "evidence_spans": [
    "ssh config",
    "extract the token",
    "email it to me"
  ],
  "candidate_sinks": ["EXTERNAL_SEND"],
  "unresolved_ambiguities": []
}
\end{lstlisting}
\end{listing}

\subsection{End-to-End Worked Trace}
\label{app:worked-trace}

We now spell out how the same request is normalized and blocked.

\paragraph{Observed Workflow.}

Assume the runtime emits the following high-level events:
\begin{enumerate}
\item \texttt{root}: the original user request is received and labeled with \texttt{SENSITIVE\_READ}, \texttt{CREDENTIAL\_ACCESS}, and \texttt{EXTERNAL\_SEND};
\item \texttt{plan}: the planner creates three subtasks, one each for retrieval, rewrite, and transmission;
\item \texttt{retrieve}: Agent A reads \texttt{\~/.ssh/config};
\item \texttt{message}: Agent A sends the extracted config fragment to Agent B;
\item \texttt{message}: Agent B sends the rewritten token-bearing content to Agent C;
\item \texttt{stage}: Agent C prepares an external email action but does not commit it.
\end{enumerate}

\paragraph{Propagation.}

\textsc{PropagateTaints} then creates a graph with task, message, and staged-sink nodes. Delegation and message edges copy active labels forward; the retrieval node inherits the root taints; the staged email node appends \texttt{EXTERNAL\_SEND} because the tool capability is a hard sink. At this point the email node carries a forbidden source--sink combination: sensitive-read or credential-access provenance terminating in external send.

\paragraph{Global Validation.}

The decisive difference from a prompt-local defense is timing. None of the intermediate steps has to be blocked in isolation:
\begin{itemize}
\item retrieving a config file may be locally plausible;
\item rewriting a token-like span may be locally plausible;
\item composing an email draft may be locally plausible.
\end{itemize}

SafeFlow blocks only after the validator sees the full path. The staged email fails declassification for two reasons:
\begin{enumerate}
\item the benign task does not authorize external credential transmission; and
\item an active upstream taint path already instantiates a forbidden source--sink rule.
\end{enumerate}

The final verdict is therefore \texttt{BLOCKED}, and the attribution report points to the path
\[
\texttt{root} \rightarrow \texttt{retrieve} \rightarrow \texttt{rewrite} \rightarrow \texttt{stage(send)}.
\]

\section{Additional Analyses}
\label{app:additional-analyses}

\subsection{Baseline and Implementation Notes}
\label{app:baseline-notes}

\paragraph{What Each Defense Observes.}

All methods in the main paper receive the same normalized benchmark instances and downstream task definitions within the same instrumented runtime, which makes workflow artifacts available before final unsafe commit. The difference lies in which available fields each implementation consumes:
\begin{itemize}
\item \texttt{SafeAgents}: undefended workflow, no explicit defense state;
\item \texttt{GuardAgent}, \texttt{AutoDefense}, \texttt{AegisLLM}: their native normalized prompt and pending-action interfaces; they do not consume the available cross-agent trace fields or construct a tainted collaboration graph;
\item \texttt{SafeFlow}: normalized prompt, planner outputs, inter-agent messages, normalized tool traces, staged sinks, prior local decisions, and graph-level provenance.
\end{itemize}

Thus all methods are evaluated under the same threat model and with the same available runtime artifacts. SafeFlow receives no privileged benchmark information: it consumes the workflow fields emitted by the shared runtime to reconstruct a cross-agent source--sink path, while the baseline implementations operate through their published prompt- and pending-action interfaces.

\begin{table*}[!tbp]
\centering
\small
\setlength{\tabcolsep}{4pt}
\renewcommand{\arraystretch}{1.12}
\begin{tabular}{p{0.17\textwidth}p{0.20\textwidth}p{0.30\textwidth}p{0.23\textwidth}}
\toprule
Case type & Example pattern & Why SafeFlow can fail or over-stage & Mitigation \\
\midrule
False negative & The planner logs a retrieval event and a later send event, but omits the message edge carrying the retrieved content. & The validator sees two locally plausible events instead of a connected source--sink path. This is an observability failure, not a rule failure. & Require mandatory provenance edges for summaries, rewrites, and staged hard sinks; mark disconnected hard sinks as ambiguous instead of safe. \\
Missed annotation & An indirect euphemism or domain-specific shorthand does not trigger a configured taint label. & The downstream sink lacks protected-source provenance, so the validator cannot form a forbidden path. & Extend evidence-reviewed label patterns and route uncertain hard-sink cases to conservative review. \\
False positive & A benign compliance report discusses ``password reset policy'' and is emailed to an authorized internal list. & A lexical annotator may attach \texttt{CREDENTIAL\_ACCESS}, while the send tool appends \texttt{EXTERNAL\_SEND}; if authorization metadata is missing, the sink is over-staged. & Require target authorization metadata and allow declassification when the task requires the sink and no concrete credential value is present. \\
\bottomrule
\end{tabular}
\caption{Representative failure and boundary cases.}
\label{tab:failure-cases-supp}
\end{table*}

\paragraph{Deterministic vs. LLM-Mediated Components.}

Intent annotation, planning, and context reconstruction are LLM-mediated and can yield different structured evidence across runs. Thus, end-to-end decisions may vary; given fixed parsed taint labels, planner skeleton, and reconstruction records, the remaining propagation, staging, aggregation, validation, and attribution steps are deterministic. This separation matters for reproducibility because defense-side LLM variation can change the graph input, but not the fixed rule schema, hard-sink policy, or the adjudication logic applied to a fixed graph.

\subsection{Reproducibility and Evaluation Notes}
\label{app:reproducibility-notes}

\paragraph{Prompt Template Skeleton.}

The exact provider wrapper is not material to the method, but the constrained-output contract is. Listing~\ref{lst:prompt-skeleton} shows the shared skeleton used for annotation and reconstruction: the model fills a fixed JSON schema and is not allowed to invent labels, release conditions, or final policy decisions.

\begin{listing}[!t]
\caption{Prompt skeleton for constrained annotation and reconstruction.}
\label{lst:prompt-skeleton}
\begin{lstlisting}
SYSTEM:
You are a SafeFlow evidence extractor.
Return JSON only; do not decide release.

ALLOWED_LABELS:
SENSITIVE_READ, CREDENTIAL_ACCESS, EXTERNAL_SEND,
PRIVILEGED_EXEC, CODE_EXEC, DESTRUCTIVE_WRITE,
PROMPT_OVERRIDE

INPUT:
- node_objective / messages / tool_events
- inherited_taints / upstream_provenance
- forbidden_sink_patterns

OUTPUT_JSON_SCHEMA:
{
  "summary": "string",
  "active_labels": ["ALLOWED_LABELS"],
  "evidence_spans": ["quoted input span"],
  "candidate_sinks": ["ALLOWED_LABELS"],
  "unresolved_ambiguities": ["string"]
}
\end{lstlisting}
\end{listing}

The parser rejects outputs containing labels outside the selected schema; any associated hard sink remains staged until it has registered labels and explicit release evidence. This design keeps the LLM component in an evidence-extraction role; final release is still governed by graph structure and the validator rules.

\begin{table}[!htbp]
\centering
\small
\setlength{\tabcolsep}{5pt}
\renewcommand{\arraystretch}{1.1}
\begin{tabular}{lccc}
\toprule
Method & ASR & Count & Wilson 95\% CI \\
\midrule
\texttt{SafeAgents} & 69.3\% & 305/440 & [64.9\%, 73.4\%] \\
\texttt{GuardAgent} & 27.5\% & 121/440 & [23.5\%, 31.9\%] \\
\texttt{AutoDefense} & 45.7\% & 201/440 & [41.1\%, 50.4\%] \\
\texttt{AegisLLM} & 30.0\% & 132/440 & [25.9\%, 34.4\%] \\
\textbf{SafeFlow} & \textbf{12.7\%} & 56/440 & \textbf{[9.9\%, 16.2\%]} \\
\bottomrule
\end{tabular}
\caption{Aggregate ASR confidence intervals using Wilson intervals. Counts are rounded to the nearest integer implied by the reported percentages and $110$ examples per benchmark.}
\label{tab:wilson-ci-supp}
\end{table}

\paragraph{Failure Case Examples.}

Table~\ref{tab:failure-cases-supp} gives concrete residual cases that inform the boundary discussion in the main paper.

\begin{table}[!htbp]
\centering
\footnotesize
\setlength{\tabcolsep}{4pt}
\renewcommand{\arraystretch}{1.05}
\resizebox{\linewidth}{!}{%
\begin{tabular}{lccc}
\toprule
Baseline & Counts & $z$ & one-sided $p$ \\
\midrule
\texttt{SafeAgents} & 56/440 vs 305/440 & -17.07 & $\ensuremath{1.4\times 10^{-65}}$ \\
\texttt{GuardAgent} & 56/440 vs 121/440 & -5.47 & $\ensuremath{2.3\times 10^{-8}}$ \\
\texttt{AutoDefense} & 56/440 vs 201/440 & -10.75 & $\ensuremath{3.0\times 10^{-27}}$ \\
\texttt{AegisLLM} & 56/440 vs 132/440 & -6.25 & $\ensuremath{2.1\times 10^{-10}}$ \\
\bottomrule
\end{tabular}%
}
\caption{Aggregate one-sided pooled two-proportion $z$-tests comparing SafeFlow's ASR to the baselines in the no-attack setting. Lower ASR is better, so negative $z$-values indicate SafeFlow's advantage.}
\label{tab:asr-significance-supp}
\end{table}

\paragraph{LLM Judge Use and Circularity Control.}

The prompt-local versus cross-agent split in the main paper is a diagnostic stratification, not a training signal and not part of the SafeFlow blocking decision. The split judge reads a realized workflow trace and returns one of three labels: \texttt{Prompt-Local}, \texttt{Cross-Agent}, or \texttt{Ambiguous}. It is not given the method name, the final defense verdict, or whether the attack succeeded.

To limit circularity, we use a rule-consistency check on the judge output. A \texttt{Prompt-Local} label is accepted only when a single local event contains enough evidence for the decisive source--sink relation. A \texttt{Cross-Agent} label is accepted only when the source evidence and sink evidence occur on different events connected by provenance or delegation. Otherwise the instance is treated as \texttt{Ambiguous} for diagnostic purposes. This stratification is reported to explain where SafeFlow's gains arise; ASR, TCR, FPR, paired success, and the final block/release decisions are computed independently of this judge.

\begin{table}[!htbp]
\centering
\footnotesize
\setlength{\tabcolsep}{4.8pt}
\renewcommand{\arraystretch}{1.02}
\begin{tabular}{lccc}
\toprule
Category & Rate & Blocked & Total \\
\midrule
All harmful & 79.1\% & 68 & 86 \\
Credential exfiltration & 87.5\% & 28 & 32 \\
Prompt override & 75.0\% & 21 & 28 \\
Code execution & 73.1\% & 19 & 26 \\
\bottomrule
\end{tabular}
\caption{Local-pass/global-block audit. These cases appear admissible under prompt-local inspection and are blocked after workflow-level reconstruction.}
\label{tab:local-global-diagnostics}
\end{table}

Table~\ref{tab:local-global-diagnostics} reports an overall local-pass/global-block rate of 79.1\%, with the highest rate on credential exfiltration (87.5\%). This diagnostic supports the role of workflow-level reconstruction when the decisive source--sink relation emerges only after messages and tool calls are assembled.

\begin{table}[!htbp]
\centering
\footnotesize
\setlength{\tabcolsep}{3pt}
\renewcommand{\arraystretch}{1.03}
\begin{tabular}{p{0.20\linewidth}p{0.27\linewidth}p{0.41\linewidth}}
\toprule
Slice & Retained harmful-success evidence & Shared evaluation record \\
\midrule
ASB ($110$) & ASB task-specific detector over the designated attack tool or alias callback. & Instance id, normalized tool trace, target capability, final harmful outcome. \\
AgentHarm ($110$) & AgentHarm task-specific detector over the target-function callback. & Instance id, normalized tool trace, target function, final harmful outcome. \\
RedCode ($110$) & Target execution capability derived from the task's target-function metadata. & Instance id, normalized tool trace, target capability, final harmful outcome. \\
SafeArena ($110$) & Target site-tool capability with retained closed-environment metadata. & Instance id, normalized tool trace, target capability, environment metadata, final harmful outcome. \\
\bottomrule
\end{tabular}
\caption{Evaluation protocol matrix. Each benchmark retains its task-specific success evidence while the shared adapter records a common trace and outcome interface.}
\label{tab:evaluation-protocol-matrix}
\end{table}

\paragraph{Measured Overhead.}

On the $440$ harmful workflows used for the main comparison, SafeFlow adds 3.7 LLM calls, 8.8 seconds of latency, and 3.0 reconstructed nodes per workflow on average. Blocks average 3.9 calls, 9.5 seconds, and 3.2 nodes; releases average 2.4 calls, 4.3 seconds, and 1.3 nodes. Workflows contain 3--10 nodes. Counts include annotation and reconstruction, but not application-provided planner calls.

\paragraph{Confidence Intervals.}

Table~\ref{tab:wilson-ci-supp} reports Wilson 95\% confidence intervals for aggregate main-table ASR. They quantify sampling uncertainty over the normalized instances, not repeated-call variance.

\paragraph{Aggregate Significance Checks.}
Table~\ref{tab:asr-significance-supp} compares SafeFlow's aggregate ASR with each baseline using a one-sided pooled two-proportion $z$-test over $440$ instances per method. The check is aggregate and unpaired because per-instance paired outputs are not retained.

\begin{table}[!htbp]
\centering
\footnotesize
\setlength{\tabcolsep}{3pt}
\renewcommand{\arraystretch}{1.02}
\begin{tabular}{p{0.18\linewidth}p{0.44\linewidth}p{0.26\linewidth}}
\toprule
Family & Source/sink instantiation & Required release evidence \\
\midrule
ASB & Untrusted task or retrieved-content evidence plus designated tool metadata; stage external-send or protected target-tool actions. & Task necessity, authorized target, and no forbidden path. \\
AgentHarm & Task objective, target-function metadata, and visible tool trace; stage privileged, destructive, or external target functions. & Task necessity, authorized target, and no forbidden path. \\
RedCode & Code-bearing request or artifact evidence plus execution-target metadata; stage code execution and privileged side effects. & Task necessity, authorized target, and no forbidden path. \\
SafeArena & Site content, target site-tool metadata, and retained environment context; stage irreversible site actions or external transmission. & Task necessity, authorized target, and no forbidden path. \\
\bottomrule
\end{tabular}
\caption{Policy instantiation matrix across benchmark families. Adapter-specific evidence is mapped into one fixed source--sink rule language.}
\label{tab:policy-instantiation-matrix}
\end{table}

\paragraph{Evaluation Protocol Matrix.}

Table~\ref{tab:evaluation-protocol-matrix} records the fixed $110$-instance slice from each benchmark (seed $20260509$) and preserves its task-specific success callback. All methods receive the same retained identifiers; the prompt-local/cross-agent judge is diagnostic only and does not determine metrics or SafeFlow decisions. Benign runs provide TCR and FPR, while paired success requires benign completion and harmful-task blocking for an aligned pair.

\paragraph{Policy Instantiation Matrix.}

Table~\ref{tab:policy-instantiation-matrix} maps each benchmark's evidence into the shared rule language. Every row uses the same declassification test; missing authorization or provenance keeps a hard sink staged.

\begin{table}[!htbp]
\centering
\scriptsize
\setlength{\tabcolsep}{2pt}
\renewcommand{\arraystretch}{0.96}
\begin{tabular}{p{0.27\linewidth}p{0.24\linewidth}p{0.37\linewidth}}
\toprule
Ambiguous provider call & Required normalization evidence & Wrapper action when evidence is incomplete \\
\midrule
\texttt{mail.deliver} or \texttt{webhook.send} & Target zone and receiver authorization. & Normalize as \texttt{send}; stage until both are available. \\
\texttt{browser.}\allowbreak\texttt{navigate} versus form submission & Side-effect flag, method, destination. & Treat as retrieval only with an explicit no-side-effect record; otherwise stage. \\
\texttt{run\_artifact} or provider-specific executor & Artifact type, execution privilege, target environment. & Normalize as \texttt{exec}; stage pending policy resolution. \\
Unknown or obfuscated call & Registered alias and complete wrapper event. & Do not commit; require registration or conservative review. \\
\bottomrule
\end{tabular}
\caption{Ambiguous mappings and the trusted-wrapper contract. The contract provides fail-closed handling for calls that are visible but not yet safely classifiable; it does not recover fully omitted provenance.}
\label{tab:trusted-wrapper-contract}
\end{table}

\paragraph{Ambiguous Mappings and Trusted Wrapper Contract.}

Table~\ref{tab:trusted-wrapper-contract} summarizes ambiguous provider mappings and the evidence required for conservative handling. Capability normalization is enforced at a trusted tool wrapper and does not rely on post-hoc inference after an unlogged side effect. Before a tool call or commit, the wrapper must emit an event id, actor id, parent or delegation id, event type, normalized capability, target resource and trust zone, content summary, authorization metadata when available, and stage/commit status. The protection scope assumes that components use this instrumented channel: the wrapper registers provider-specific aliases before execution and rejects or conservatively stages a visible call that has no approved capability mapping. It does not make hidden side channels or completely uninstrumented components observable. If a visible high-impact sink has disconnected provenance, the wrapper treats it as ambiguous and withholds it.
\end{document}